# Unpacking uncertainty in the modelling process for energy policy making

Samuele Lo Piano[a),*], Máté János Lőrincz[a)], Arnald Puy[b),c)], Steve Pye[d)], Andrea Saltelli[e)],

Stefán Thor Smith[a)], Jeroen P. van der Sluijs[c),f)]

a) School of the Built Environment, University of Reading, UK
b) Department of Ecology and Evolutionary Biology, Princeton University, New Jersey USA
c) Centre for the Study of the Sciences and the Humanities (SVT), University of Bergen, Norway
d) UCL Energy Institute, University College London, UK
e) Open Evidence, Universitat Oberta de Catalunya, Spain
f) Department of Chemistry, University of Bergen, Norway

* s.lopiano@reading.ac.uk

# Highlights

- We discuss critical issues in modelling for energy policy making, including model non-neutrality and uncertainty

- We propose the NUSAP scheme, diagnostic diagrams, and sensitivity auditing as responsible approaches to these critical issues

- We showcase the application of these approaches using practical examples

- Joint uptake of these tools could benefit the overall policy-making process

# Abstract

This paper explores how the modelling of energy systems may lead to undue closure of alternatives by generating an excess of certainty around some of the possible policy options. We exemplify the problem with two cases: first, the International Institute for Applied Systems Analysis (IIASA) global modelling in the 1980s; and second, the modelling activity undertaken in support of the construction of a radioactive waste repository at Yucca Mountain (Nevada, USA). We discuss different methodologies for quality assessment that may help remedy this issue, which include NUSAP (Numeral Unit Spread Assessment Pedigree), diagnostic diagrams, and sensitivity auditing. We demonstrate the potential of these reflexive modelling practices in energy policy making with four additional cases:

(i) stakeholders' evaluation of the assessment of the external costs of a potential large-scale nuclear accident in Belgium in the context of the ExternE (External Costs of Energy) project;

(ii) the case of the ESME (Energy System Modelling Environment) for the creation of UK energy policy;

(iii) the NETs (Negative Emission Technologies) uptake in Integrated Assessment Models (IAMs); and

(iv) the Ecological Footprint (EF) indicator. We encourage modellers to widely adopt these approaches to achieve more robust and inclusive modelling activities in the field of energy modelling.

# List of acronyms

BECCS: Bioenergy with Carbon Capture and Storage

EC: European Commission

ECC: External Cost Calculation

EF: Ecological Footprint

ESME: Energy System Modelling Environment

IAMs: Integrated Assessment Models

IIASA: International Institute for Applied Systems Analysis

IPCC: International Panel on Climate Change

NDCs: Nationally Determined Contributions

NETs: Negative Emission Technologies

NPP: Nuclear Power Plant

NUSAP: Numerical Unit Spread Assessment Pedigree

TSPA: Total System Performance Assessment

UNFCCC: United Nations Framework Convention on Climate Change

# 1. Introduction

Models are representations of a system. In this very feature resides their power: encoding the relations of the system that is represented and decoding the relations to understand the properties of the system from the model (Rosen 1991). Energy system modelling has become a mainstay of many policy processes, having been used extensively in the UK since 2003, for example, to inform the different energy and climate strategies, as well as legislative target-setting (Williams et al. 2018). However, energy policy is informed and often led by small sets of models and system representations. This is the case, for instance, in the energy policy making of the European Union, which has been extensively based only on the use of the findings of the PRIMES (Price-Induced Market Equilibrium System) model (Clark 2011; European Commission 2019). The methodological approach, or technical stance, adopted in a modelling activity is not neutral, however; it actually conditions the narrative produced within an analysis and consequently the decision-making it is meant to inform (Saltelli, Benini, et al. 2020). This aspect can lead to important controversies in a decision-making setting, especially when not transparently disclosed.

On top of this, the use of quantification has significantly increased over the last decades with the inflation of metrics, indicators, and scores to rank and benchmark options. The case of energy policy making in the European Union is again an effective example. The European Union's recent energy strategy has been underpinned by the *Clean Energy for all Europeans* packages, which are in turn underpinned by a number of individual directives, each one characterised by a series of quantitative goals (European Commission 2017). The quantification of the impact is usually required to successfully promote new political measures (European Commission, 2015). This impact assessment is also based on modelling activities. The emphasis on producing exact figures to assess the contribution of a new technology or political or economic measure has put many models and their users into contexts of decision-making that have gone beyond their original intent.

In their quest for capturing the features of the energy systems represented, models have increased their complexity. In this context, the need to appraise model uncertainty has become of paramount importance, especially considering the growing uncertainty arising due to propagation errors caused by model complexification. In ecology this is known as the O'Neil conjecture which posit a principle of decreasing returns for model complexity when uncertainties come to dominate the output (O'Neill 1989; Turner and Gardner 2015). Capturing and apportioning uncertainty is crucial for a healthy interaction at the science-policy interface, including energy policy making, because it promotes more aware decision-making. Yet Yue et al. (2018) found that only about 5% of the studies covering energy system optimisation models have included some form of assessment of stochastic uncertainty, which is the part of uncertainty that can be fully quantified (Walker et al. 2003). When it comes to adequately apportioning this uncertainty onto the input parameters and hypotheses through sensitivity analysis, the situation is even more critical: only very few papers in the energy field have developed state-of-the-art approaches (Saltelli et al. 2019). Furthermore, the epistemic part of uncertainty, which is the part that arises due to imperfect knowledge and that is beyond full quantification, has been practically ignored in the vast majority of the energy modelling literature (Pye et al. 2018).

In this contribution, we propose approaches to deal with the challenges of non-neutrality and uncertainty in models. These challenges are especially critical when only one (set of) model(s) has been selected to inform decision-making. We then illustrate how these issues may come into play in two practical cases: first, the International Institute for Applied Systems Analysis (IIASA) global modelling in the 1980s; and second, the modelling activity in support of the construction of a radioactive waste repository at Yucca Mountain.

In turn, we discuss approaches to deal with the challenges of uncertainty and scope for decision-making in modelling. The Numerical Unit Spread Assessment Pedigree (NUSAP) method, diagnostic diagrams, and sensitivity auditing are the methodologies we present. We highlight the application of the methods proposed to deal with the issue identified by reviewing four seminal case studies: ESME (Energy System Modelling Environment) and the estimate of the external cost of a nuclear accident, UK energy modelling for policy-making support, the uptake of NETs in IAMs, and the Ecological Footprint (EF). Finally, we draw conclusions on the lessons learned and their implications for policy-making.

## 2. Methods

The approaches we discuss have been inspired by the epistemological reflexivity of post-normal science (PNS). PNS becomes relevant in the presence of disputed values, high stakes, urgent decisions, and uncertain facts, all aspects that characterise energy policy making (Funtowicz and Ravetz 1993). This characterisation of energy policy is justified by factors such as trade-offs of socio-environmental aspects; possible massive and threatening environmental impacts; and unknown relations and causalities in the socio-technical and environmental domain. Policy-making adds to this an unavoidable dimension of complexity.

We selected NUSAP, diagnostic diagrams, and sensitivity auditing for this contribution due to their capacity to address crucial aspects of the assumptions and modelling relations underpinning quantification. The appraisal of the assumptions is of paramount importance at the science-policy interface, a fortiori in a controversial policy domain such the energy field.

The NUSAP system for communication and management of uncertainty assesses the broader dimensions of uncertainty in quantitative analysis (Funtowicz and Ravetz 1990; van der Sluijs 2017; van der Sluijs et al. 2005). This approach retains the strengths of quantitative uncertainty assessment but adds a focus on the assessment of the quality or 'pedigree' of the underlying model assumptions. It broadens the critical appraisal of knowledge to include several dimensions and locations of uncertainty in the modelling approach, including model structure (relationships embedded in equations), and model inputs, including data, system boundaries, and problem frames (Petersen et al. 2013; van der Sluijs and Petersen 2008; Walker et al. 2003).

Pedigree is judged against multiple criteria based on a structured scoring system (van der Sluijs, Risbey, and Ravetz 2005). These criteria include

- the *proxy* representation of the real-world system (how good or close a measure of the modelled quantity is to the actual quantity represented),
- the *empirical basis* of the numbers used (the degree to which direct observations, measurements, and statistics are used to estimate the parameter),
- the *rigour* of the methods to derive numbers (the norms for methodological rigour in this process applied by peers in the relevant disciplines),
- *validation* (the degree to which one has been able to cross-check the data and assumptions used to produce the value of the parameter against independent sources), and
- level of *theoretical understanding* of the systems being modelled (the extent and partiality of the theoretical understanding that was used to generate the value of that parameter).

The pedigree scores can be combined with the quantitative indices obtained from a sensitivity analysis on how the output uncertainty is affected by a given input/assumption. These two dimensions of uncertainty – stochastic (sensitivity indices) and epistemic (pedigree score) – can be visualised in relation to each other using a *diagnostic diagram*. However, capturing the quantitative part of uncertainty in a technical sensitivity analysis does not suffice to address the epistemic background and scope of a quantification. To this end, Saltelli et al. (2013) sought to enhance sensitivity analysis with sensitivity auditing, which has also been recently taken as one of the core ingredients of a manifesto for responsible quantification (Saltelli, Bammer, et al. 2020). Sensitivity auditing is based on a seven-point checklist that allows for a systematic check of the scope and background of a quantification, including modelling activities and indicators, in a policy-making context. The checklist covers the following points:

- **Rhetorical use:** Check against a rhetorical use of mathematics – Are large models being used where simpler ones would suffice?
- **Assumption hunting:** What assumptions were made? Were these explicit or implicit?
- **Detect GIGO:** Detect garbage in, garbage out (GIGO) – Was the uncertainty in the input artificially constrained to boost the model's certainty? Or, conversely, was it bloated so as to, for example, prevent regulators from making decisions in a case of harmful products?
- **Anticipate criticism:** Find sensitive assumptions before they find you – It is better to anticipate criticism by undertaking robust uncertainty and sensitivity analyses before publishing one's results.
- **Aim for transparency:** Black box models do not play well in a public debate.

- **Do the right sum:** Do the right sums, not just the sums right – Is the issue properly identified or does the model address the 'wrong' problem? Or, is the model addressing a closed definition of what the problem might be instead of including multiple perspectives/stakeholders?

- **Perform UA and SA:** Perform thorough and state-of-the-art uncertainty and sensitivity analyses.

Sensitivity auditing has been recommended in guidelines for impact assessment, including those of the European Commission (2015). The Science Advice for Policy by European Academies (2019) also recommended the use of both NUSAP and sensitivity auditing.

# 3. Case studies

In this section, we discuss the issues introduced in energy policy making along with the approaches proposed to tackle them in practical case studies. The full list of cases is in Table 1.

**Table 1.** Case studies discussed in this paper.

| Limitations in energy-related modelling | | |
|---|---|---|
| **Case study** | **Energy policy issue** | **Method** |
| International Institute of Applied Analysis (IIASA) global modelling | Political stance of the modelling activity implemented, uncertainty | - |
| Modelling in support of Yucca Mountain radioactive waste repository | Uncertainty, scope of the model for decision-making | - |
| **Improved approaches to modelling energy-related matters** | | |
| **Case study** | **Energy policy issue** | **Method** |
| External cost Estimation for nuclear energy | Epistemic uncertainty | Pedigree analysis for NUSAP |

| Energy modelling support for UK policy | Stochastic and epistemic uncertainty | NUSAP and diagnostic diagram |
| --- | --- | --- |
| Negative emission technologies uptake in IAMs | Stochastic and epistemic uncertainty | NUSAP and (adapted) diagnostic diagram |
| Ecological footprint accounting | Uncertainty, scoping of the model for decision-making | Sensitivity auditing |

## 3.1 Limitations in energy-related modelling: the cases of IIASA and Yucca Mountain

### 3.1.1 International Institute for Applied Systems Analysis (IIASA) global modelling

The goal of the IIASA's Energy Project in the 1980s was to "understand the factual basis of the energy problem, i.e., to identify the facts and conditions for any energy policy" (Häfele et al. , 1981). Table 2 summarises the project's breadth and the magnitude of the IIASA energy modelling. The European Commission used the IIASA scenarios to develop a community energy policy, which also plays an important role in national decision-making committees (Häfele 1981).

**Table 2.** Overview of IIASA Energy System Program (Source: Häfele et al. ,1981).

| Duration | 7 years |
| --- | --- |
| Number of researchers involved | More than 140 |
| Background of the research team group members | Economics, Physics, Engineering, Geology, Mathematics, Psychology, Psychiatry, and Ethnology |

| Number of countries involved | 19 (from developed and developing countries) |
|---|---|
| Number of partners | 35 |
| Institution involved | United Nations Environment Programme, International Atomic Energy Agency, National Center for Atmospheric Research, Electric Power Research Institute, Stanford Research Institute, Nuclear Research Center Karlsruhe, Institut Economique et Juridique de l'Energie, Volkswagen Foundation, Federal Ministry of Research and Technology, Meteorological Office, National Coal Board, Austrian National Bank, and Siberian Power Institute |
| Number of research reports and conference proceedings | 60 |

The IIASA energy model was built from a collection of interconnected sub-models:

- MEDEE-2, an accounting framework-based energy demand model developed by the University of Grenoble.

- MESSAGE, a dynamic linear programming-based energy supply and conversion system model developed at IIASA.

- IMPACT, an input-output model for calculating the impacts of alternative energy scenario origins from the Siberian Power Institute.

- MACRO, a macroeconomic model developed in Canada and the USA.

- An oil trade gaming model produced by the Siberian Power Institute (Häfele et al. 1981; Keepin 1984).

The IIASA's set of energy models is shown in Figure 1, along with the most relevant inkages among the many parts composing the system. Most of the feedback is provided by manual calculations: changes to one set of inputs do not automatically propagate across models (Basile, 1980).

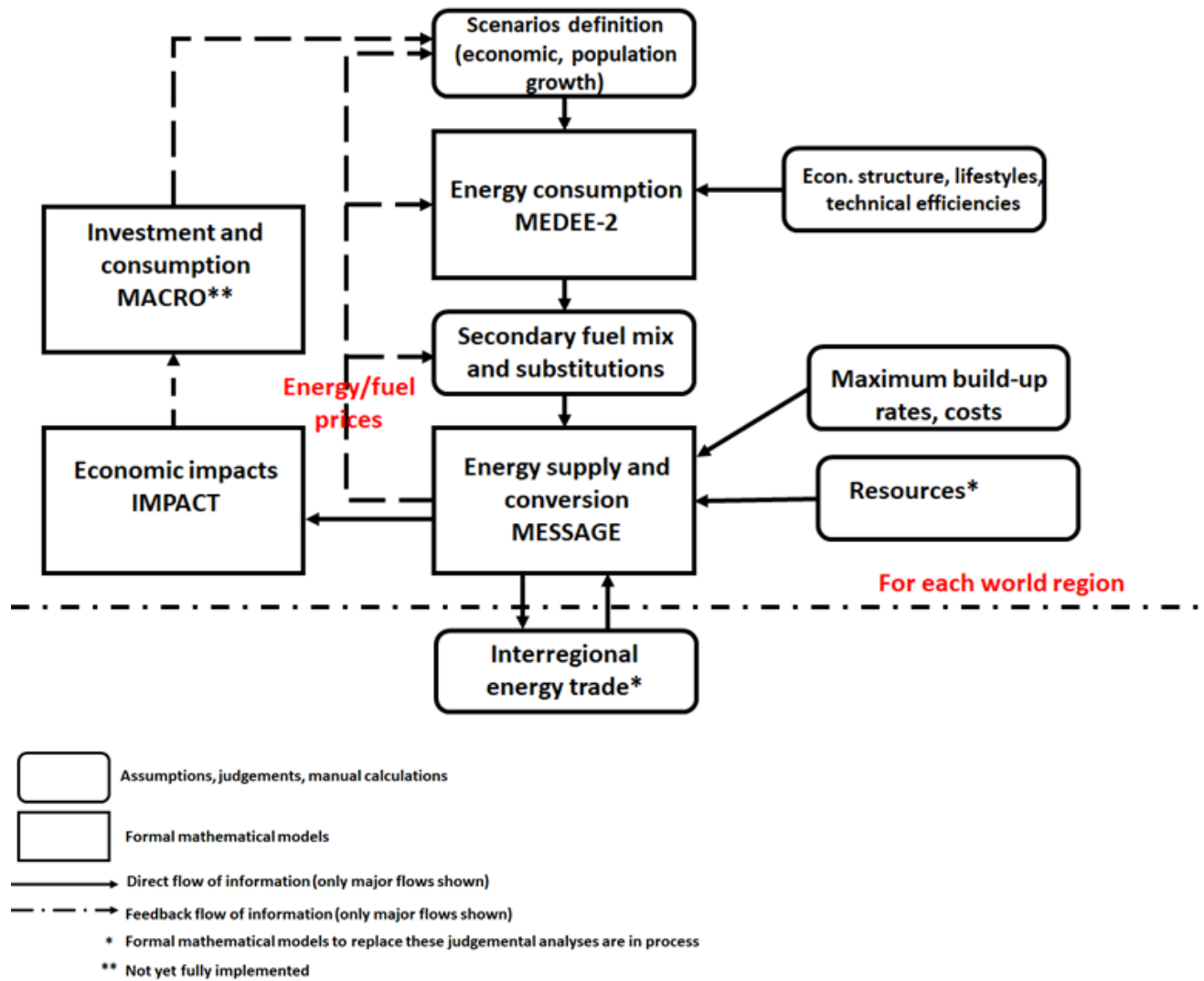

**Figure 1.** The IIASA's set of energy models (based on Basile 1980, p. 6).

Among the significant problems encountered in this approach was the failure of the feedback link between IMPACT and MEDEE-2. This meant that economic variables such as costs and resource extraction limits were not incorporated into the iteration process (Wynne 1984). IMPACT was involved in only a few of the major 'iterations' and was directed by a range of informal judgments (for example, about future capital/output ratios).

This indicates that the models did not account for capital, land, labour, technology, water, and material investment. The basic assumptions of the IIASA analysis were that the global population and average per capita income would have doubled by 2030, which implied that a sustainable future would only be possible through the expansion of all energy resources required for sustainability. Thereby, Häfele (1976) primarily focused on large-scale nuclear technology in developing future energy strategies as per the following criteria:

- $CO_2$ emissions reduction on a global scale by gradually substituting nuclear energy for fossil fuels.
- Maximum utilization of finite nuclear fuels via closed-loop recirculation and breeder technology, resulting in a virtually infinite source of energy.
- Diversification of resources with all energy sources within their optimal operating ranges.
- Secure energy supply with the energy mix base, reserve, and buffer capacity.
- The use and integration of 'small-scale technologies' into the power grid with an ever-present nuclear power.
- Total $CO_2$ avoidance and substitution of fossil fuels utilizing coal gasification, methanation, and hydrogen technology.

In the end, the main conclusion of the IIASA's Energy Project was that the transition to fast-breeder reactors and large-scale solar and 'coal synfuels' had to be made, and could be achieved by 2030 if these power plants were promptly and largely deployed (Keepin 1984). In the IIASA projections, the share of nuclear power was estimated to rise to 77% in 2030. Keepin (1984) showed in a sensitivity test that minor changes to assumptions produced high effects on the results, which contradicted the "robust conclusions" of this modelling activity (Figure 1). An alternative scenario was developed by this author showing that an increase of nuclear costs by 16%, and of the coal extraction limit by 7%, would have resulted in phasing out the nuclear path.

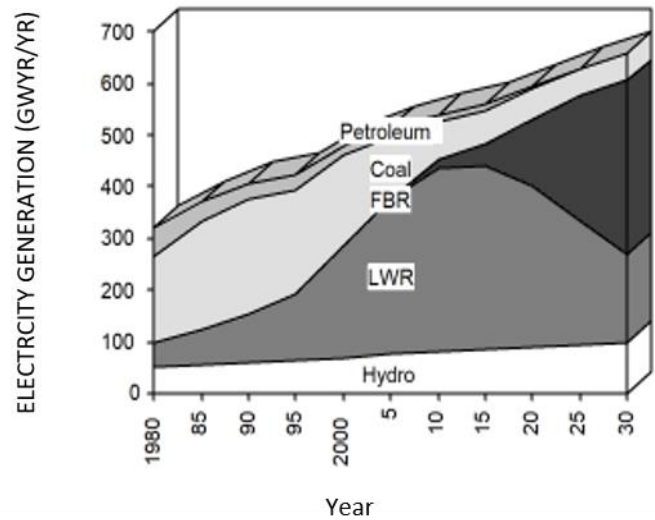

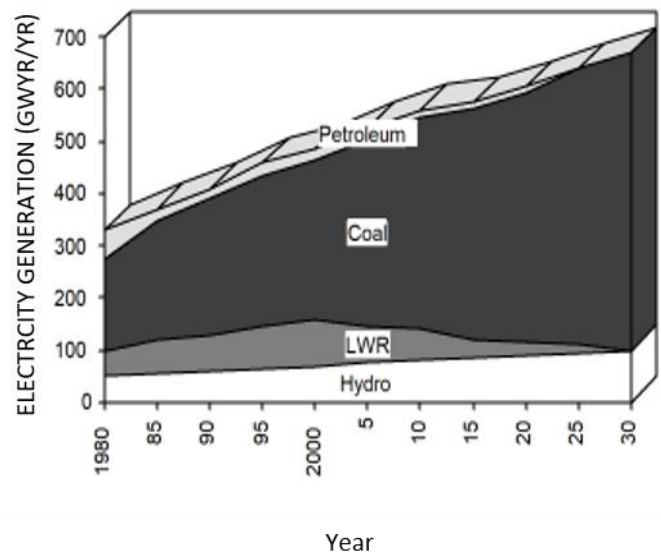

**Figure 2.** Sensitivity to cost assumptions in the IIASA energy demand scenario for the United States and Canada (LWR = Light Water Reactors; FBR = Fast Breeder Reactors). 'Coal' includes both present and advanced technologies. Original scenario results for electricity generation (source: Middtun et al. 1986, reproduced with permission).

On an epistemic level, the IIASA modellers adopted from relational ecology the concept of resilience (Holling, 1973). The concept of resilience was first proposed by Holling in the field of ecology, who demonstrated that complex natural systems developed distinct strategies to respond to perturbations such as inevitable environmental change. To compute the shifting boundaries of points of equilibrium in the IIASA modelling activity, a reliable method for measuring potential fields and basins of attraction was required. Häfele saw resilience as a way to boost the credibility of his scenarios (Häfele, 1976). He attempted to replicate Holling's resilience by dividing it into smaller 'resilience basins'. However, Holling opposed the use of the resilience concept in nuclear energy due to the high risk and uncertainty of the technology entailed. As a result, Holling dismissed the Häfele team's proposals for massive nuclear parks, energy islands, and hydrogen pipelines as 'bad science fiction'. This forced one's desired option into another's concept, without recognising the loss of information inherent in the process. (Hulme 2011) defines this as epistemological slippage. The authors also note that one aspect of resilience is the avoidance of lock-ins and path-dependencies. On this score resilience does not appear to favour nuclear energy.

The modelling exercise performed by IIASA is an illustration of how modelling can be used to influence the the trajectories of technology (Keepin, 1984). Overall, the issue with this type of policy-oriented modelling is the lack of quality control and of proper interdisciplinary scrutiny.

This IIASA modelling activity needs to be understood in the context of the 1970s energy crisis. Energy policy required accessible science, which IIASA set out to deliver. However, this case study shows that building a convincing modelling activity may require broadening the perspective from relying solely on one class of scientific experts; this could favourably open up and expose the social and institutional assumptions embedded in the modelling activity.

### 3.1.2 The Yucca Mountain radioactive waste repository

IIASA modelling has contributed to the uptake of nuclear energy, which is predicated on the safe disposal of nuclear waste. The following questions, therefore, become crucial: Can the safety commitment of the relevant industrial and institutional actors be guaranteed over time for the safe disposal of nuclear waste? Can modelling help in this respect?

This section explores the case of the construction of a nuclear waste deposit within the Yucca Mountain, Nevada.

The operation of civil nuclear energy and its societal acceptance relies on the safety of nuclear waste disposal due to the hazrd posed by nuclear waste. As a result, waste confinement should be guaranteed for a long period of time. After an initial period of considering options such as deep bed oceanic disposal or firing waste-loaded spacecrafts towards the sun, experts converged on the option of geological disposal (Saltelli et al., 1988). In the US, the Yucca Mountain, a tuff formation in the Nevada desert, was chosen for experimentation and possibly as an ultimate disposal site.

The risk assessment of the Yucca Mountain site was based on the use of a model that described nuclide migration from container to human interaction through the following stages:

    (i)      Release from container vessels due to estimated corrosion rates of containers

    (ii)     Transport through buffer material surrounding containers

    (iii)    Transport through local geosphere

    (iv)    Selected pathways for human exposure, such as ingestion and inhalation

The Total System Performance Assessment (TSPA) software, developed by the US Department of Energy and comprising 286 sub-models with thousands of parameters, was mandated to explore the safety of the Yucca Mountain disposal site. Originally, planners sought assurances that the Yucca Mountain site would be safe for 10,000 years; this was later increased to one million years under a federal court ruling based on the long half-life of certain radionuclides (Pilkey and Pilkey-Jarvis 2009).

One of the main assumptions of TSPA was later challenged by the site investigation. While the model predicted long times for the water to reach the repository level, the discovery of $^{36}$Cl (a man-made radioactive isotope of chlorine) at the level of the repository, 300 meters underground, contested this assumption. The $^{36}$Cl isotope was associated with the fallout of nuclear weapons tests carried out before 1963, and its presence implied that water could travel 300 metres downward through the matrix of the rock in less than 50 years. According to the percolation rate used in TSPA (less than 10 mm per year), this time should have been of the order of tens of thousands of years. This result was based on the assumption that the tuff composing the geological formation could be modelled as a homogeneous medium without preferential pathways for the migration of contaminants. For a time, it was hence considered that the disposal operation was not meeting the required standards. However, recent determinations have questioned the presence of 'pulse isotopes' at the repository (Cizdziel and Smiecinski 2006; Stothoff and Walter 2013), thus leaving the entire matter still hanging on political decisions (Sarewitz 2018).

The modelling of the disposal system relied on a single large model to describe the physics, geology, chemistry, and biology of the system. This approach gave prominence to an agenda of defined, manageable uncertainties. What it neglected, however, were the less reducible uncertainties of the associated institutional and social setting. Wynne (1992) discusses these settings using the concept of indeterminacy, as distinguished from uncertainty. Indeterminacy is linked to the answers to questions such as: Will the high-quality maintenance, inspection, and operation of a risky technology be sustained in the future, multiplied over replications, and possibly with many instances all over the world?

If one contrasts the mathematical precision of TSPA with the confusion and neglect surrounding the present-day's nuclear waste in dump sites in the US (Diaz-Maurin 2018), it is easy to grasp the relevance of the point made by Wynne.

In particular, the modelling activity around TSPA, which aimed to predict the fate of radionuclides thousands of years into the future,[1] has never apparently stopped (Helton, Hansen, and Swift 2014). The point made in this section is that the model used, TSPA, has locked the policy discussion of the fate of nuclear waste into one where safety can be proven over geological times, which is clearly an impossible task.

### 3.2 Improved approaches to modelling energy-related matters

### 3.2.1. External costs of nuclear energy: a pedigree analysis of assumptions

Across Europe, debates about the place of nuclear power production in a sustainable energy mix have resulted in enduring and intractable conflicts between actors holding antagonistic positions. External cost calculations (ECC) can be seen as a particular form of evidence that can inform decision-making regarding sustainable energy. The use of ECCs, the data used to underpin them, and their use in policy making is, however, controversial because of the complexity of the policy issue and the plural positions, values, and stakes that are at play. To critically review ECCs and their use in informing policy-making, the Belgian nuclear research centre organized a transdisciplinary workshop with policy makers, stakeholders, and experts from various disciplines. This was undertaken in the context of the European ExternE project (1990-2005) that aimed to quantify the environmental external costs of different energy supply options (EC, 1995; 1999). The case is documented in detail in Craye, Laes, & van der Sluijs, (2009) and Laes, Meskens, & van der Sluijs (2011), a brief summary of which is presented here. The workshop applied the concept of 'pedigree of knowledge' first, after which crucial assumptions and choices made in the calculation chain were identified, critically appraised, and qualified in a structured way amongst all workshop participants. The discussion focused on assumptions related to the scenario used to evaluate the impacts of a severe nuclear accident in a Belgian Nuclear Power Plant (NPP) near Antwerp, including the estimation of the related health impacts and their economic and monetary valuation. Special attention was paid to the so-called value-laden character of these assumptions, following the method proposed by Kloprogge et al., (2011).

---

[1] In the interest of disclosure, one of the authors of the present work (AS) has computed the effects of radiological doses on a population up to one hundred thousand years into the future (http://www.oecd-nea.org/tools/abstract/detail/nea-0860/).

Preparatory interviews yielded a gross list of 30 assumptions in the ECCs. By means of an internet survey involving all workshop participants, a final list of the six most critical assumptions was selected for further appraisal in the workshop:

> (i) External costs of a potential large-scale accident in a Belgian NPP can be determined based on a calculation for a hypothetical NPP located in the middle of Western Europe.

(ii) In a large-scale accident scenario for a Belgian NPP, all radionuclide dispersion routes other than the atmospheric release route are negligible.

(iii) A linear correlation exists between exposure to ionising radiation and health effects, even for very small radiation doses.

(iv) All health impacts other than the radiological ones caused by exposure to ionising radiation can be neglected when assessing the consequences of a large-scale nuclear accident in a Belgian NPP.

(v) The 'risk-aversion factor' for accidents of the 'low probability/ high consequences' type cannot be reliably determined, and therefore does not have to be reported.

(vi) The cost indicators adopted in the ExternE methodology (i.e., cost of countermeasures, direct economic damage, and short- and long-term health impacts) are sufficiently representative for the total costs of a potential large-scale nuclear accident in a Belgian NPP.

Overall, the six assumptions received weak pedigree scores. The panel generally considered these assumptions not very plausible, subject to disagreement, and, to a large extent, inspired by contextual factors rather than grounded in evidence. The only exception was the third assumption of linear correlation. Expert participants agreed that the so-called 'linear no-threshold hypothesis' constituted the best scientific basis for regulating the risks of ionising radiation and served as a precautionary approach to managing radiation risks. The assessment also resulted in clear suggestions for improvements of the ECCs as regards assumptions (i) and (ii). Assumption (i) was contested as the specific location of the Belgian NPPs (near major cities with important industrial activities) was more likely to result in higher externalities in comparison with the hypothetical location of mid-Western Europe that was used in the assessment. Criticism of assumption (ii) was motivated by the impossibility to exclude contamination of ground and river water by radionuclides in the case of a severe nuclear reactor accident. Such contamination could, for instance, occur in case of a failure of the NPP's pressure vessel or after a melting of the reactor core through the bottom of the reactor building, resulting in steam explosions. This highly improbable accident sequence would result in long-term ecosystem pollution, which was excluded as an externality in the ExternE approach because it only considers impacts on the human health count.

A second category of assumptions consisted of those that were more severely criticized and for which only limited proposals could be made to improve the ECCs. For instance, in the case of assumption (vi), many types of economic impacts were excluded from the ExternE methodology. These included direct and indirect costs of lost production in industries adjacent to the NPP, forward ripple effects in the entire European economy (e.g., caused by an evacuation of the Antwerp harbour), costs of 'stigmatization' of a region contaminated by nuclear fallout, and economic impacts on the nuclear sector worldwide (e.g., costs of cancelling new nuclear programmes and enhanced safety measures in existing plants). The discussion of this assumption led most of the participants to agree to a high likelihood for continued public contestation of ECCs that systematically exclude such costs.

Some assumptions were seen as problematic, but it was unclear how to overcome the related problems through other calculation methods. As regards assumption (iv), one expert participant argued that follow-up studies from the Chernobyl reactor accident in 1986 showed a significant increase in the population suffering from issues such as post-traumatic stress symptoms, anxiety, estrangement, and dislocation. However, it was difficult to relate these psychological impacts unequivocally to the nuclear accident alone because the risk management interventions of the Soviet authorities would have also played a role. Therefore, non-radiological (psychosomatic) health impacts of a potential reactor accident could prove to be a major— albeit hardly quantifiable—factor.

Engaging stakeholders in a NUSAP-based workshop made it possible to hold a transparent debate in which the assumptions underpinning the ESME modelling activity were openly scrutinized. This provided valuable input for the negotiation aimed at improving the ESME modelling process.

3.2.2 UK energy modelling as a support to policy making

In the UK, scenario analysis using energy models has often suffered from deterministic thinking. A view of uncertainty has been adopted using a storyline-and-simulation approach. Practitioners have increasingly realised that this is insufficient when dealing with complex and uncertain transition that must happen over relatively short timescales (Usher and Strachan 2012). Based on ex-post analysis of modelled energy futures, this narrow approach has been shown to be limited, with real-world developments occurring that are completely outside of the anticipated range (Craig, Gadgil, and Koomey 2002; Trutnevyte et al., 2016). As a result, modelling practice in the UK has (to some extent) evolved and shifted towards a range of quantitative approaches to dealing with uncertainty, from probabilistic analysis ( Pye, Sabio, and Strachan 2015; Pye, Usher, and Strachan 2014) through to stochastic programming (Usher and Strachan 2012) and modelling-to-generate-alternatives (Li and Trutnevyte 2017; Trutnevyte 2016). There is also a recognition that the Government requires more information on uncertainty, as outlined in its Aqua Book on analytical quality assurance (HM Treasury 2015).

While these approaches push in the right direction, by their very nature of being quantitative they are likely to overlook uncertainties that are not easily quantifiable (van

der Sluijs et al., 2005). These include the strength of the underlying knowledge base underpinning the modelling, or the degree to which the many assumptions made by modellers are value-laden.

To broaden the assessment of uncertainties in energy modelling, the NUSAP approach was applied to a UK-based modelling exercise. This concerned ESME, a key model used for research informing UK government on energy issues (McGlade et al. ,2018; S. Pye, Sabio, and Strachan 2015). The approach to the exercise was based on the following steps: (i) identify assumptions that materially affect the model results, through global sensitivity analysis and expert elicitation; (ii) determine criteria against which to assess pedigree; (iii) run the stakeholder workshop to generate the scores; and (iv) compare pedigree results to quantitative model results using a diagnostic diagram (Pye et al. ,2018).

In the ESME exercise, a global sensitivity analysis was used to determine which input parameters the model solution was most sensitive to. In other words, if a policy maker is looking for a cost-effective strategy, the sensitivity analysis seeks to identify the input assumptions that had the greatest influence on the costs of that strategy. The influence of different factors (representing quantitative uncertainty) was then plotted against the pedigree scores of those same assumptions using a diagnostic diagram (van der Sluijs et al., 2005).

Figure 3 indicates that some of the technology assumptions that are important for UK energy and climate policy have a weak pedigree. These land in Q4, a quadrant termed the 'danger zone', where assumptions have high sensitivity scores but weak pedigree. In this category belong bioenergy resource assumptions (BioRES), which is crucial for biofuels in sectors such as international aviation and BECCS (Bioenergy with Carbon Capture and Storage), and CCS deployment (CCSmbr), which is again important for BECCS but also for hard-to-mitigate sectors such as iron and steel and cement. The value of such information for decision-makers is that they should proceed with caution when drawing policy conclusions from model solutions that rely heavily on bioenergy and CCS.

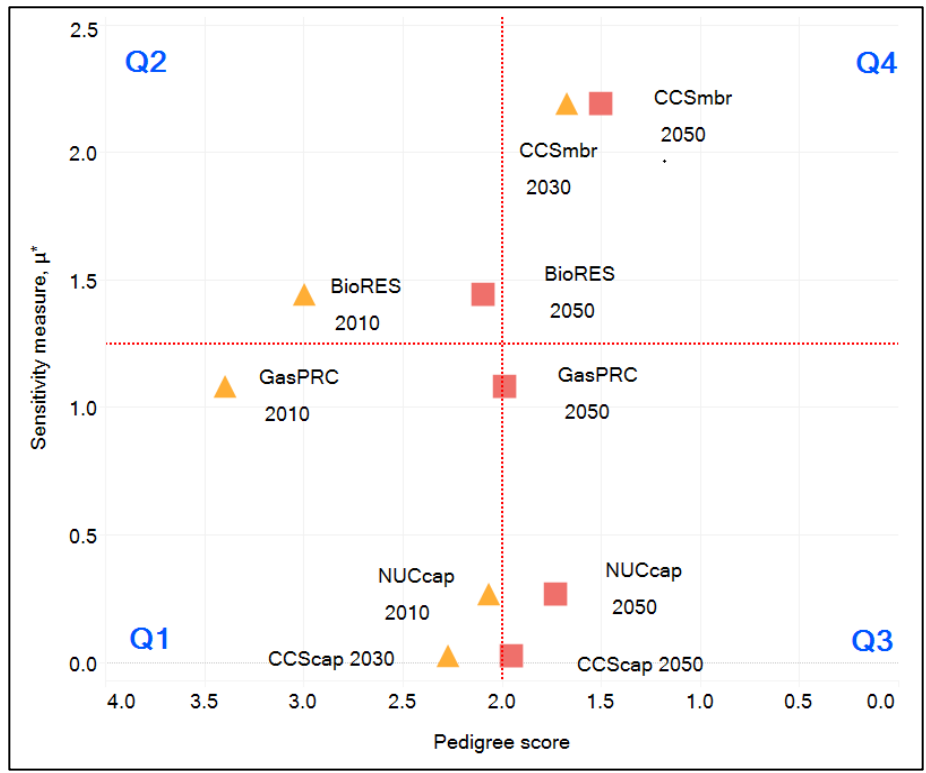

**Figure 3. Diagnostic diagram to compare qualitative (pedigree scores, horizontal axis) against quantitative (sensitivity measure, vertical axis) uncertainties.** The sensitivity measure (based on the elementary effects method (Saltelli et al., 2008) or Morris method (Morris 1991)) highlights the influence of the modelled uncertainty on the variance across the model objective function, which is the total discounted system costs.

In addition to pedigree scoring, stakeholders were asked to score assumptions in terms of the extent to which an assumption was *justifiable and defensible*, and whether a specific assumption would likely find *agreement amongst peers*. The results highlighted that there are often many reasonable and possible choices for different assumptions. This emphasises the need for transparency around modelling choices and a debate on, and scrutiny of, assumptions with broad stakeholder input.

Beyond the modelling insights from the exercise, the process of the workshop was equally valuable. It allowed expanding the range of stakeholders, while providing time to explain the logic behind critical model assumptions. It also demonstrated how to better engender trust in the analytical process and broaden expert input into the exercise. Given that previous modelling has been viewed as a black box, the process itself can enable critical scrutiny and contribute to the process of making energy systems analysis more transparent for decision-makers (Cao et al. ,2016; Pfenninger 2017).

### 3.2.3 Negative emission technologies

Routes to meeting the targets of the 2015 Paris Agreement imply a commitment to reduce anthropogenic greenhouse gas (GHG) emissions. To achieve this goal, two broad energy-technology approaches are considered: (i) real reduction of emissions from renewable energy technology (primary focus); and (ii) negative emission technology (NET) as abatement of continued emissions. The use of NETs falls under a broader category of geoengineering, that are debated due to the potential unforeseen consequences to the environment, as well as reducing the collective commitment of society to environmental sustainability (Sovacool 2021).

NETs can make the requirements regarding emission cuts less stringent by enhancing the planetary $CO_2$ sink capacity (van Vuuren et al., 2017). For instance, reductions of anthropogenic $CO_2$ emissions have been estimated in IAM activities at 60-85% or 70-95% for 2050, relative to 2010 figures, dependent on whether BECCS (a type of NET) is being deployed or not (van Vuuren et al. ,2017). Indeed, BECCS plays a prominent role in the NET literature (Anderson and Peters 2016), to the point that large uptakes of BECCS have been posited in IPCC IAMs scenarios. These have been questioned in the literature on the basis of the following: first, there being too few existing plants (Babin, Vaneeckhaute, and Iliuta 2021); second, the capability of delivering negative emissions over the time span of the cultivations (Hanssen et al., 2020); third, the effectiveness of the coupled afforestation/re-forestation strategies (Turner et al., 2018, Krause et al., 2018); fourth, the important amount of land required (Field & Mach (2017); and finally, the pace in increased land cultivation required to contribute to meeting the climate goals set for 2100 (Turner et al., 2018). Such a massive implementation could conflict with other fundamental sustainability goals such as food security and biodiversity conservation (Dooley, Christoff, and Nicholas 2018).

The general picture that emerges from this criticism is that BECCS deployment would seemingly entail substantial stakes in return for benefits that are very uncertain. Presenting the outcome of these models with crisp figures generates the exact result of leaving out other potential options.

If uncertainties were acknowledged, other options would become comparable and worth investigation.

As regards the case of BECCS uptake in IAMs, Workman et al. (2020) identified the following issues:

(i) the certainty on key assumptions (such as feasibility, cost, and deployment rates) over several decades was overestimated;

(ii) values beyond monetary proxies were excluded; and

(iii) linked to the previous point, representatives of a single community define goals for climate policy rather than having these resulting from a dialogue among multiple stakeholders.

Butnar et al. (2020) added criticisms on the scarce transparency of IAMs models relative to modelling assumptions, as well as on the treatment of the socio-cultural and institutional dimension.

Public acceptance could be facilitated by opening up the IAMs modelling activity to multi-criteria assessments. These are capable of including values beyond monetary proxies used in cost optimisations (Stephens et al., 2021; Workman et al., 2020). However, this would require significant effort to update these modelling activities, their scope, and their theoretical background (Braunreiter et al., 2021).

Tavoni et al. (2017) investigated the potentially problematic nature of the underpinning assumptions of NETs uptake in IAMs. Based on a consultation with experts from the field, these authors identified as particularly resistant to modelling dimensions of governance, public acceptance, external costs, and impacts. By contrast, modelling was more propitious to appreciate the importance of operational costs and effectiveness.

In this highly-debated context, Vaughan & Gough (2016) resorted to the NUSAP method by engaging 18 experts in a workshop to scrutinise several key assumptions fed into IAM models. The identified 9 key assumptions related to bioenergy (available land area, future yield, and proportion of energy); CCS (storage capacity, technology uptake, capture rate); and cross-cutting assumptions (policy framework, social acceptability, and net negative emissions). The authors also made use of a diagnostic diagram whereby the pedigree score was assessed against a qualitatively estimated influence on results through a dedicated pedigree score. Even in this case, similar to the case study on ESME, most of the discussed assumptions ended up in the danger zone of high influence on the result coupled to a weak pedigree score.

This case study illustrates how an issue in modelling activity identified in the literature can be brought to the fore and discussed with relevant peers. The process enables mutual learning, while placing under the spotlight potential criticalities in the modelling activity.

### 3.2.4 Ecological footprint

Ecological footprint (EF) is a successful sustainability indicator proposed by the Global Footprint Network. Diverse sources have advocated for its use as an indicator to lead energy policy making (Abbas, Kousar, and Pervaiz 2021; Metcalf 2003). Energy consumption accounts for the most important part of the EF measure (Giampietro & Saltelli 2014). This part is bound to increase in the future because of BECCS deployment and further land allocation for energy uses. EF measures human demand on natural capital, which is understood as the quantity of natural land (expressed in global hectares equivalents) required to support an individual or economic activity. The 'Earth overshoot day' is the date by which humanity will have used all available natural resources from the Earth's yearly natural budget. The systematic anticipation of Earth overshoot day over the years is widely recognised as a sign of humanity's unsustainable pattern of economic development (Giampietro and Saltelli 2014b). We make use here of the seven-point sensitivity auditing checklist, presented in the methods section, to evaluate whether EF is an adequate indicator to capture this aspect.

- Rhetorical use: According to Giampietro and Saltelli (2014), the ecological footprint has been systematically over-interpreted in terms of representing the planet's biocapacity. What is presented in the EF as a measure of what can be produced within the planet's ecological limits is merely an accounting of agricultural productivity. Several other dimensions are excluded from EF accounting as per other points examined below.

- Assumption hunting: A potentially misleading feature in EF accounting concerns Its bioenergy dimension. For instance, the question of how the $CO_2$ absorbing capacity decreases with the ageing of forests is neglected. The same caveat applies to the paradox that replacing natural ecosystems with more productive human-made vegetation would lead to an improvement of the planet's biocapacity rather than to an impoverishment due to loss of biodiversity and natural habitats (Giampietro and Saltelli 2014).

- Detect GIGO: Several potential sources of uncertainty remain unaddressed in EF accounting. No error in terms of biocapacity is considered, nor is the variable of accuracy discussed at the local, national, and global levels. A data quality score is the only proxy included at the country level. This leads to an issue in terms of how the information is processed and aggregated across scales, as examined in more detail below.

- Anticipate criticism: The rounding of values and the cascading of uncertainty across scales is one of the factors contributing to the fragility of EF accounting. To the best of our knowledge, this uncertainty has not been accounted for, let alone apportioned through sensitivity analysis, in the modelling adopted in EF accounting.

- Aim for transparency: The documentation on EF accounting is available, but some technical coefficients are not openly traceable. This is the case for the equivalence factors, which reflect the relative productivity of world average hectares of different types of land use. How these quantities were arrived at can only be retrieved from a satellite workbook, which is available only upon request.

- Do the right sum: The EF accounting does not help in defining whether types of land allocation actually contribute to sustainability (Galli et al., 2016) . One example is that of a landfill, which is crucial on the waste sink side, but whose importance is entirely missed in the EF biocapacity accounting.
- Perform UA and SA: As previously discussed, uncertainty in the accounting is largely overlooked, with the exception of a data quality score as proxy at the country level. Hence, the space of assumptions has not been explored at all, which leaves unaddressed the responsiveness of the EF indicator to its uncertainty sources.

The usefulness of sensitivity auditing stems from its capacity to highlight the limitations of a quantification, in this case the EF, when using it for policy making in real-world cases.

# 4. Conclusions and policy implications

A quantification, mathematical model, or any large system of indicators can be thought of as a cathedral, an ancient *fabrica*, which is never finished; new bits are added or modified over time, bugs are solved, and new questions are posed. In this construction, choices are made all the time; these choices may concern the use of a physical or heuristic law, the value of variables, or which algorithm to choose among the many available to solve a problem. We use the word 'choice', or 'assumption', as more than one possible item or value could be selected, but eventually just one enters the model construction. It is only normal that with time, not even its developers will be in condition to remember all that was chosen. This sedimentation of modelling assumptions, which enables the model to answer the questions asked from it, also constitutes an obstacle to its transparency. The only way out of this predicament is for modellers to make many choices, many assumptions, and propagate them through the model. Instead of a prediction, of a single point in the multidimensional space of model outputs, we now have a cloud.

This process of retracing one's steps in order to rediscover the forgotten choices and assumptions, and to perform the analysis just described – which would appear technically as an uncertainty quantification – is facilitated if one incorporates in the analysis the philosophies of NUSAP and sensitivity auditing that we illustrated in this contribution. Additionally, when engaging stakeholders, interfacing the two approaches in a diagnostic diagram can offer a thorough view of the uncertainty at play, as shown in this contribution's case study of UK energy modelling for support to policy making. It is also noteworthy that this process of going back in order to go forward plays an important role in system thinking (Koestler 1989). This process, the modelling of the modelling process, is also valuable in unearthing path dependences and lock-ins, to identify stages where a given issue became frozen in a dead-alley configuration of conflict among stakeholders. Going forward, in turn, may take the form of a broadening of the spectrum of the policy options.

In the present work, we have seen how several issues may impact quantification in energy systems in producing scenarios, indicators, and modelling activities. Untested assumptions and implicit political stances may result in assessments that are not 'politically robust' (i.e., cannot be shared by all stakeholders), as shown in the case of the IIASA modelling activity with the controversy around proposing to largely resort to nuclear power as a means to achieve resilient energy system.

Given the importance of uncertainty assessment in modelling for public policy, the degree to which model assumptions are value-laden and have 'pedigree' are dimensions of uncertainty that need to be considered by the community of modellers. Engaging stakeholders in workshops offers the chance to explore the logic, perspective, and framework that lies at the base of critical model assumptions, to demonstrate how to better engender trust in the analytical process, and to broaden expert input into the exercise (see the external cost of nuclear energy and the UK energy policy making case studies). The renegotiation of the assumptions that inspired a move towards quantification can take place through direct interaction between the involved stakeholders, relevant experts, and policy makers, ideally in a setting that allows experimentation with the socio-institutional roles 'normally' entrusted to restricted communities of experts/regulators at the science–policy–society interface. All these aspects emerged from the ESME modelling activity and the assessment of the external cost of nuclear energy, in which the group of experts/stakeholders/policy makers engaged in the workshop put under scrutiny six key modelling assumption and put forth concrete suggestions for their improvement. The application of sensitivity auditing to the ecological footprint indicator highlighted its most questionable aspects in the capacity to capture useful metrics for sustainability, hence contributing to potentially setting ill-conceived policy goals. The objection that these approaches are costly or encourage paralysis by analysis are unfounded for the case of energy policy, given the important stakes and long time-horizons covered by the policy. The applications of these methods to currently active topics such as the modelling of NETs in IAMs appears also promising. Having lawmakers demand the use of the analytical lenses suggested in the present work would help to strengthen the quantifications that underpin energy policy making, leading to tangible benefits for the overall policy-making process.

# Conflicts of interest

The authors declare that they have no known competing financial interests or personal relationships that could have appeared to influence the work reported in this paper.

# Funding

This research has received funding from UK Research and Innovation grant agreement EP/R035288/1 as part of the Centre for Research into Energy Demand Solutions. Arnald Puy worked on this paper under a Marie Sklodowska-Curie Global Fellowship (grant number 792178).